\documentclass{article}

\usepackage[english]{babel}

\usepackage[letterpaper,top=2cm,bottom=2cm,left=3cm,right=3cm,marginparwidth=1.75cm]{geometry}

\usepackage{amsmath}
\usepackage{graphicx}
\usepackage[colorlinks=true, allcolors=blue]{hyperref}
\usepackage{amssymb}
\usepackage{float}
\usepackage{listings}
\usepackage[english]{babel}
\usepackage{amsmath, amssymb, amsfonts}
\usepackage{graphicx}
\usepackage{hyperref}
\usepackage{caption}
\usepackage{listings}
\usepackage{xcolor}
\usepackage{booktabs}
\usepackage{float}
\usepackage{amsmath}
\usepackage{listings}
\usepackage{xcolor}
\usepackage{cite}

\title{Eigen-Portfolios: From Single-Component Models to Ensemble Approaches}
\author{Zhengxiang Zhou, Yuqi Luan}
\date{}

\begin{document}
\maketitle

\begin{abstract}
The increasing integration of data science techniques into quantitative finance has enabled more systematic and data-driven approaches to portfolio construction. This paper investigates the use of Principal Component Analysis (PCA) in constructing eigen-portfolios---portfolios derived from the principal components of the asset return correlation matrix. We begin by formalizing the mathematical underpinnings of eigen-portfolios and demonstrate how PCA can reveal latent orthogonal factors driving market behavior. Using the 30 constituent stocks of the Dow Jones Industrial Average (DJIA) from 2020 onward, we conduct an empirical analysis to evaluate the in-sample and out-of-sample performance of eigen-portfolios. Our results highlight that selecting a single eigen-portfolio based on in-sample Sharpe ratio often leads to significant overfitting and poor generalization. In response, we propose an ensemble strategy that combines multiple top-performing eigen-portfolios. This ensemble method substantially improves out-of-sample performance and exceeds benchmark returns in terms of Sharpe ratio, offering a practical and interpretable alternative to conventional portfolio construction methods.
\end{abstract}

\section{Introduction}

The intersection of machine learning and financial modeling has led to a surge in interest around data-driven investment strategies. Among these techniques, \cite{hotelling1933pca} and \cite{tipping1999probabilistic} show that Principal Component Analysis (PCA) has gained prominence for its ability to reduce dimensionality and uncover latent structure within high-dimensional financial datasets. In the context of portfolio management, PCA enables the decomposition of asset return correlations into uncorrelated components\cite{guo2018eigenportfolio}, which can then be interpreted as synthetic portfolios—termed \textit{eigen-portfolios}—each corresponding to a distinct source of market variation.

This paper focuses on the practical construction and evaluation of eigen-portfolios using historical return data from the Dow Jones Industrial Average (DJIA). Specifically, we explore whether portfolios constructed from individual or combined principal components can achieve superior risk-adjusted performance compared to standard benchmarks.

The study is structured in two major parts. The first part presents the theoretical foundation, including the standardization of return data, eigenvalue decomposition of the correlation matrix, and the derivation of eigen-portfolios. In the second part, we empirically assess the performance of these portfolios, identifying the principal component that maximizes in-sample Sharpe ratio and analyzing its stability in out-of-sample testing. To address the observed overfitting and instability, we further propose an ensemble method, which is introduced in \cite{dorabiala2023ensemble}, that combines multiple top-ranking eigen-portfolios, offering improved generalization and robustness.

This investigation contributes to the literature by providing both a methodological framework and empirical evidence for the benefits and limitations of PCA-based portfolio construction. Our results suggest that while PCA can reveal meaningful risk factors, naive implementation is prone to instability, and ensemble methods are better suited for practical portfolio management in evolving market environments.

\section{Mathematical Foundations of the Eigen-portfolio}

\subsection{Standardized Return Matrix}

Let \( P \in \mathbb{R}^{T \times N} \) denote the matrix of asset prices, where each column corresponds to a different asset and each row represents a different time point.

The \textbf{daily linear returns} are computed using the first-order percentage change:

\[
R_{t,i} = \frac{P_{t,i} - P_{t-1,i}}{P_{t-1,i}}, \quad \text{for } t = 2, \dots, T
\]

This yields a return matrix \( R \in \mathbb{R}^{(T-1) \times N} \), where each element \( R_{t,i} \) is the return of asset \( i \) on day \( t \).

Before applying Principal Component Analysis (PCA), all variables (i.e., asset return series) must be placed on a comparable scale. If not standardized, assets with higher return variance will dominate the principal components due to their greater numerical magnitude.

To achieve this, each column (asset) in the return matrix is standardized to have zero mean and unit variance. For each asset \( i \), the standardized return \( \tilde{R}_{t,i} \) is given by:

\[
\tilde{R}_{t,i} = \frac{R_{t,i} - \mu_i}{\sigma_i}
\]

where:
\begin{align*}
\mu_i &= \frac{1}{T-1} \sum_{t=1}^{T-1} R_{t,i} \quad \text{(mean return of asset } i\text{)} \\
\sigma_i &= \sqrt{ \frac{1}{T-2} \sum_{t=1}^{T-1} (R_{t,i} - \mu_i)^2 } \quad \text{(standard deviation of asset } i\text{)}
\end{align*}

This transformation maps the return distribution of each asset to a standard normal distribution:

\[
\tilde{R}_{t,i} \sim \mathcal{N}(0, 1)
\]

Remark: Standardization ensures that the covariance or correlation matrix used in PCA reflects structure in the data, not scale differences between variables.

\subsection{Correlation Matrix}

Using the standardized return matrix \( \tilde{R} \), we compute the empirical \textbf{correlation matrix} \( \rho \in \mathbb{R}^{N \times N} \) as:
\[
\rho = \frac{1}{T-1} \tilde{R}^\top \tilde{R}
\]
Since \( \tilde{R} \) is standardized, \( \rho \) is equivalent to the correlation matrix of the original returns.

\subsection{Eigenvalue Decomposition}

We perform eigenvalue decomposition of the correlation matrix:
\[
\rho = Q \Lambda Q^\top
\]
where:
\begin{itemize}
    \item \( Q = [q_1, q_2, \dots, q_N] \in \mathbb{R}^{N \times N} \) is a matrix whose columns are orthonormal eigenvectors,
    \item \( \Lambda = \mathrm{diag}(\lambda_1, \lambda_2, \dots, \lambda_N) \) is a diagonal matrix of eigenvalues,
    \item \( \lambda_1 \geq \lambda_2 \geq \dots \geq \lambda_N \geq 0 \).
\end{itemize}

Each eigenvector \( q_i \) defines a synthetic, uncorrelated portfolio — the \textbf{eigen-portfolio} — corresponding to a principal component of asset return variation.

\subsection{Eigen-portfolios}

The weights of the \( i \)-th eigenportfolio are given by the normalized eigenvector:
\[
\pi_i = \frac{q_i}{\sum_{j=1}^{N} q_{ij}}
\]
so that the portfolio weights sum to one:
\[
\sum_{j=1}^{N} \pi_{ij} = 1
\]

Remark: The variance of the eigenportfolio returns is equal to the associated eigenvalue:
\[
\mathrm{Var}(\pi_i^\top \tilde{R}_t) = \lambda_i
\]

\subsection{Cumulative Explained Variance}

To assess how many eigen-portfolios are needed to explain the majority of variation in the asset returns, we compute the \textbf{cumulative explained variance} (CEV):
\[
\mathrm{CEV}(k) = \frac{\sum_{i=1}^k \lambda_i}{\sum_{i=1}^N \lambda_i}
\]

Remark: Typically, a small number of eigenportfolios explain a large portion of the variance, indicating the presence of strong latent factors.

\subsection{Return Reconstruction}

The standardized returns can be approximated using the first \( k \) eigenvectors:
\[
\tilde{R}_t \approx \sum_{i=1}^{k} \alpha_{ti} q_i
\]
where:
\[
\alpha_{ti} = \tilde{R}_t q_i
\]
is the projection of the standardized return vector at time \( t \) onto the \( i \)-th eigenvector. This provides a dimensionality-reduced representation of asset returns.

\section{Construction of an eigen-portfolio}

\subsection{Basic Idea}

The construction of an eigen-portfolio is grounded in principal component analysis (PCA) applied to the returns of the 30 constituent stocks of the Dow Jones Industrial Average (DJIA). The procedure begins by forming a standardized return matrix from the historical price data of these stocks. The dataset is then split chronologically, with the first 80\% of observations used as a training set and the remaining 20\% reserved for out-of-sample testing.

PCA is performed on the training set to obtain the principal components, which are the eigenvectors of the sample covariance (or correlation) matrix. Each principal component is a 30-dimensional vector representing a linear combination of the stocks—effectively, a set of portfolio weights.

Each eigenvector is normalized such that the absolute weights sum to one, or optionally scaled to sum to 100, to reflect percentage allocations. These vectors are treated as candidate portfolios, which we evaluate by applying them to the training set to compute their historical Sharpe ratios.

The eigen-portfolio corresponding to the highest Sharpe ratio on the training data is selected. This optimal portfolio is then applied to the testing set to assess its out-of-sample performance \cite{gustrigos2024eigenportfolio}.

\subsection{Exploratory Data Analysis}

Principal Component Analysis (PCA) is most effective when applied to datasets with significant correlation among variables, as it leverages this correlation structure to reduce dimensionality\cite{uwpca2024}. So We start by examining the correlation between each stock.

\begin{figure}[H]
    \centering
    \includegraphics[width=0.7\textwidth]{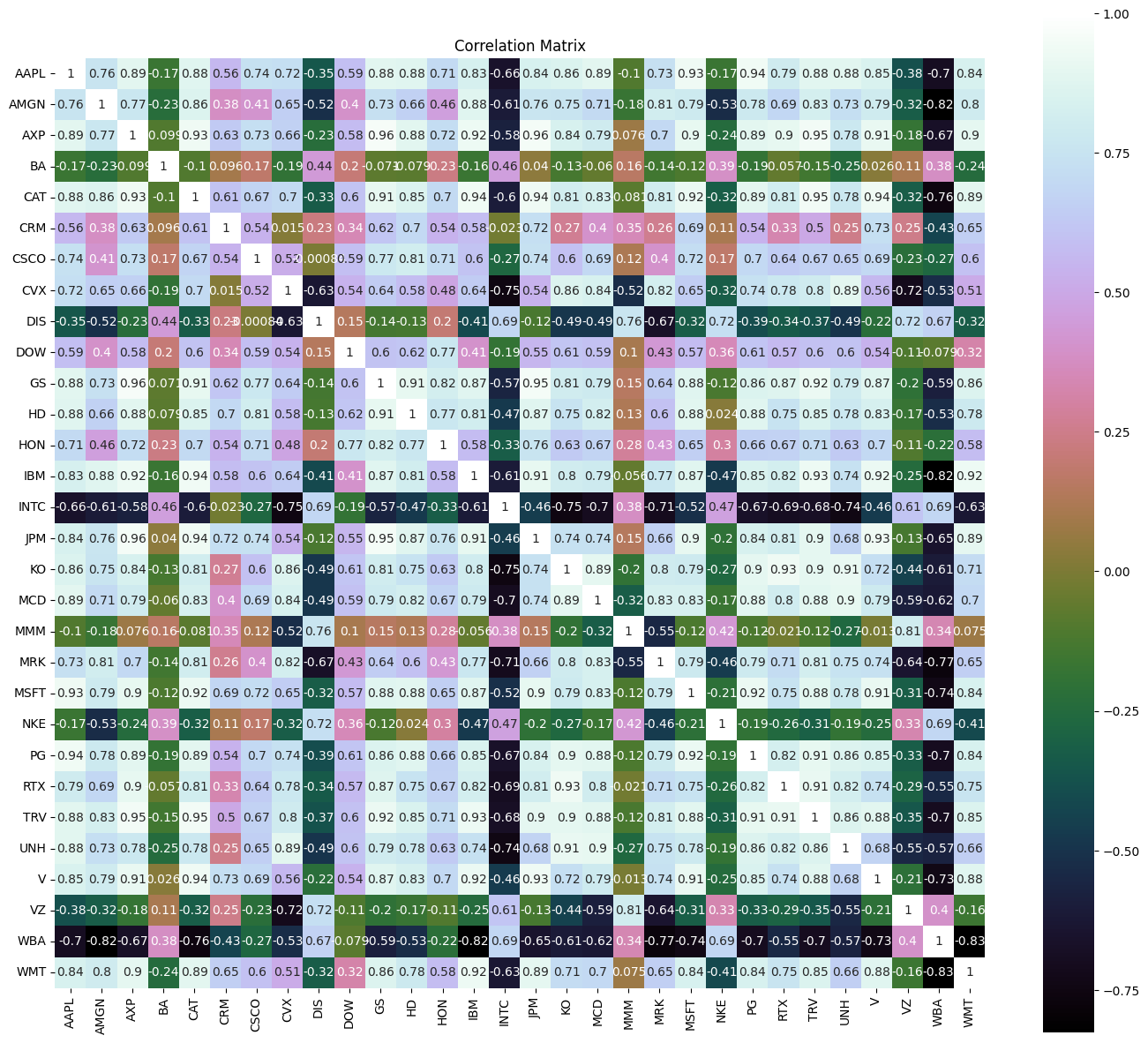}
    \caption{Colored correlation plot of the 30 constituent stocks of the Dow Jones Industrial Average (DJIA). In this context, preliminary analysis indicates a high degree of correlation among the assets. So it suggests that the return dynamics of these stocks are influenced by common underlying factors.}
    \label{fig:corr_plot}
\end{figure}

Given this structure, PCA is a suitable technique for extracting these latent factors and constructing eigen-portfolios. By identifying principal components that explain the majority of the variance, we can potentially represent the behavior of the entire portfolio using a smaller number of orthogonal factors. This not only aids in dimensionality reduction but also enhances interpretability and portfolio construction efficiency.

\vspace{1\baselineskip}

Next, we analyze the explained variance derived from Principal Component Analysis (PCA).

\begin{figure}[H]
    \centering
    \includegraphics[width=0.8\textwidth]{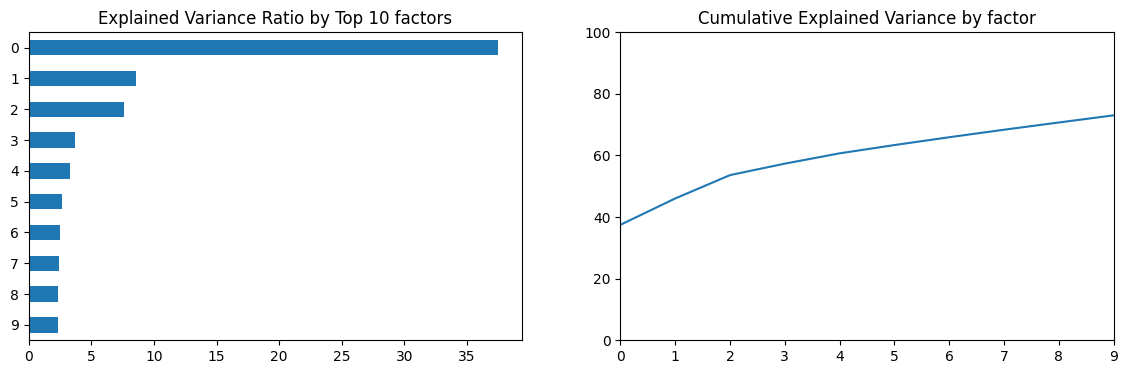}
    \caption{Explained variance of PCA components. The first principal component captures approximately 37\% of the total daily return variation across the asset universe, and is often interpreted as representing a broad market factor. The cumulative explained variance plot indicates that the top 10 components together account for about 73\% of the total variance, suggesting a relatively low-dimensional structure underlying the cross-section of stock returns.}
    \label{fig:explained_var_plot}
\end{figure}

\vspace{1\baselineskip}

To better understand the composition of the eigen-portfolios derived from PCA, we examine the portfolio weights associated with the principal components. Each principal component corresponds to a vector of loadings (weights) across the 30 DJIA stocks, which can be interpreted as the allocation of a hypothetical portfolio that loads entirely on that component.

The figure below presents bar charts of the portfolio weights for the first five principal components, which are ranked by the amount of variance they explain in the standardized return matrix. These components can be viewed as uncorrelated latent factors driving stock returns.

Each subplot represents one eigen-portfolio. The height of each bar indicates the weight of a given stock in that portfolio. A positive weight implies a long position, while a negative weight represents a short position.

\begin{figure}[H]
    \centering
    \includegraphics[width=0.8\textwidth]{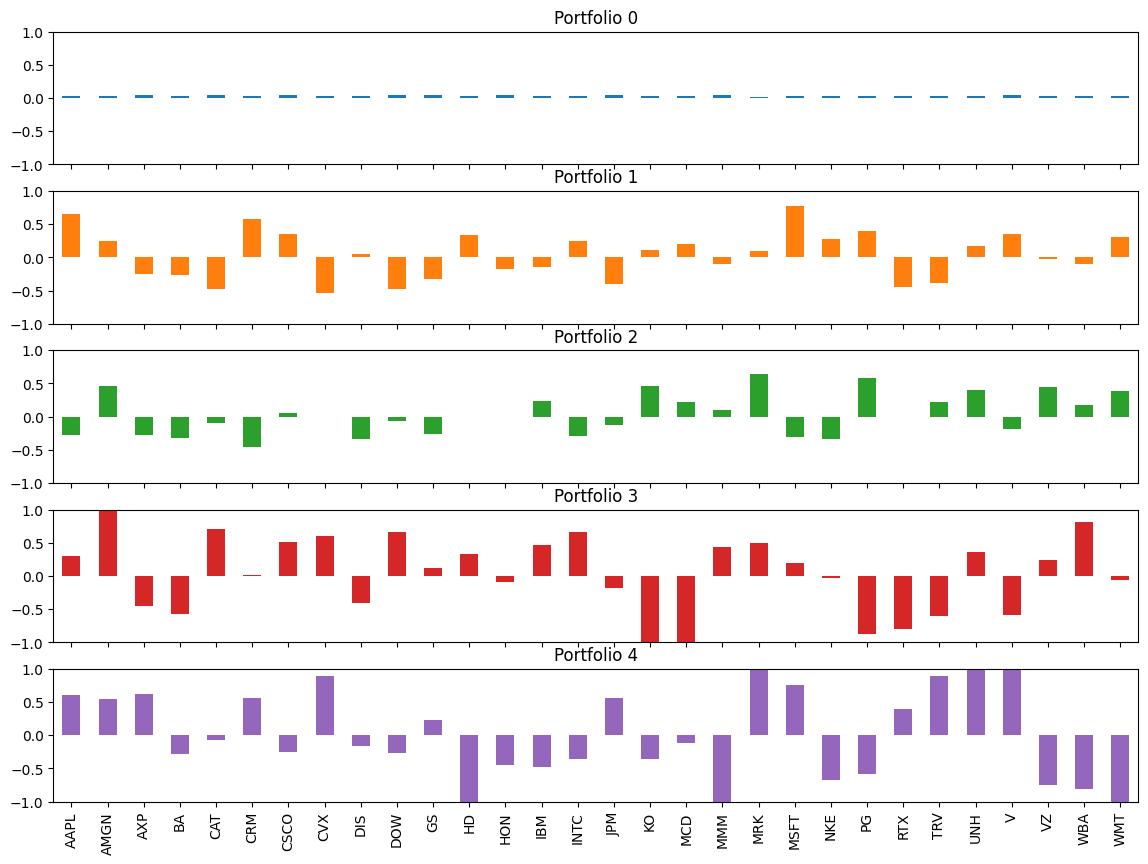}
    \caption{Portfolio weights associated with the first five principal components derived from PCA. Each bar corresponds to one of the 30 DJIA stocks. The sign and magnitude of each weight reflect the relative exposure of the eigen-portfolio to that stock.} 
    \label{fig:sample_portfolio_weight}
\end{figure}

\subsection{Identifying the Optimal Eigen-Portfolio on the Training Set}

To identify the most promising eigen-portfolio for backtesting, we employ the Sharpe ratio—a widely used performance metric that quantifies risk-adjusted returns. Specifically, the Sharpe ratio is calculated as the ratio of a portfolio’s annualized return to its annualized volatility. A higher Sharpe ratio indicates better performance, either through higher returns, lower risk, or both.

The annualized return is computed using the geometric mean of the portfolio’s periodic returns, scaled according to the number of trading periods in a year (e.g., 252 for daily returns). Volatility is measured as the standard deviation of returns, annualized by multiplying by the square root of the number of periods per year.

We iterate through all principal components derived from PCA, treat each eigenvector as a set of portfolio weights, and apply the Sharpe ratio function to evaluate performance. The component with the highest Sharpe ratio is selected as the optimal eigen-portfolio on the training set.

\begin{figure}[H]
    \centering
    \includegraphics[width=0.7\textwidth]{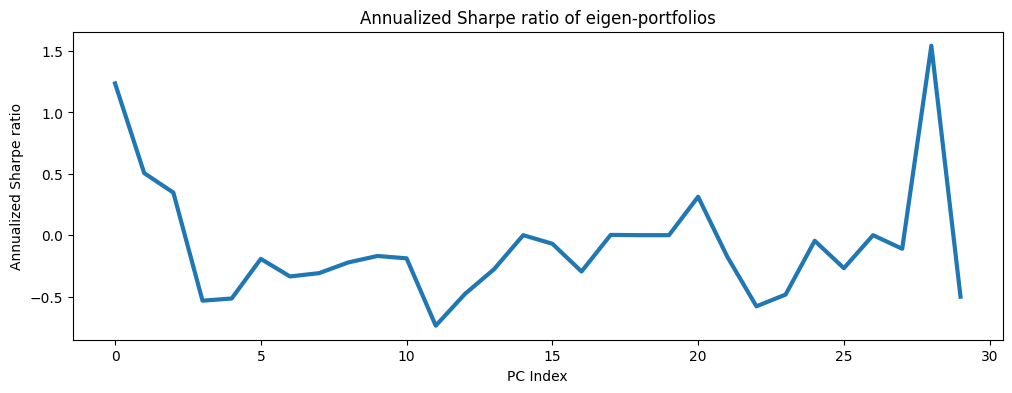}
    \caption{Shows that Principal Component 28 achieves the highest Sharpe ratio.} 
    \label{ei train}
\end{figure}

\begin{table}[H]
    \centering
    \begin{tabular}{r|r|r|r}
        \textbf{PC Index} & \textbf{Annualized Return} & \textbf{Annualized Volatility} & \textbf{Sharpe Ratio} \\
        \hline
        28 & 2.434612 & 1.576678 & 1.544140 \\
        0  & 0.176736 & 0.142872 & 1.237029 \\
        1  & 0.202403 & 0.400229 & 0.505717 \\
        2  & 0.157184 & 0.452778 & 0.347156 \\
        20 & 0.338361 & 1.080644 & 0.313110 \\
        17 & 0.001927 & 0.869588 & 0.002216 \\
        24 & -1.000000 & 21.857259 & -0.045751 \\
        15 & -1.000000 & 14.337079 & -0.069749 \\
        27 & -0.271841 & 2.452422 & -0.110846 \\
        9  & -0.228341 & 1.346807 & -0.169543 \\
        21 & -0.421951 & 2.404992 & -0.175448 \\
        10 & -0.999998 & 5.298901 & -0.188718 \\
        5  & -0.133552 & 0.691459 & -0.193146 \\
        8  & -0.999412 & 4.495682 & -0.222305 \\
        25 & -0.804200 & 2.987966 & -0.269146 \\
        13 & -0.992816 & 3.578139 & -0.277467 \\
        16 & -0.999438 & 3.376408 & -0.296006 \\
        7  & -0.208852 & 0.675296 & -0.309275 \\
        6  & -0.196401 & 0.583769 & -0.336437 \\
        12 & -0.792039 & 1.651738 & -0.479519 \\
    \end{tabular}
    \caption{Clear demonstration of Sharpe ratios, annualized returns, and volatilities for selected eigen-portfolios computed from the training set. Principal Component 28 achieves the highest Sharpe ratio of 1.54, corresponding to an annualized return of approximately 243\%, suggesting strong in-sample performance.}
    \label{tab:sharpe_train}
\end{table}

\subsection{Examining the Optimal Eigen-Portfolio}

Having identified the eigen-portfolio with the highest Sharpe ratio on the training set, we now examine its composition in greater detail. Specifically, we analyze the portfolio weights assigned to each of the 30 constituent stocks of the DJIA. This provides insight into which stocks contribute most significantly—either positively or negatively—to the portfolio’s overall risk-adjusted performance.

\begin{figure}[H]
    \centering
    \includegraphics[width=0.7\textwidth]{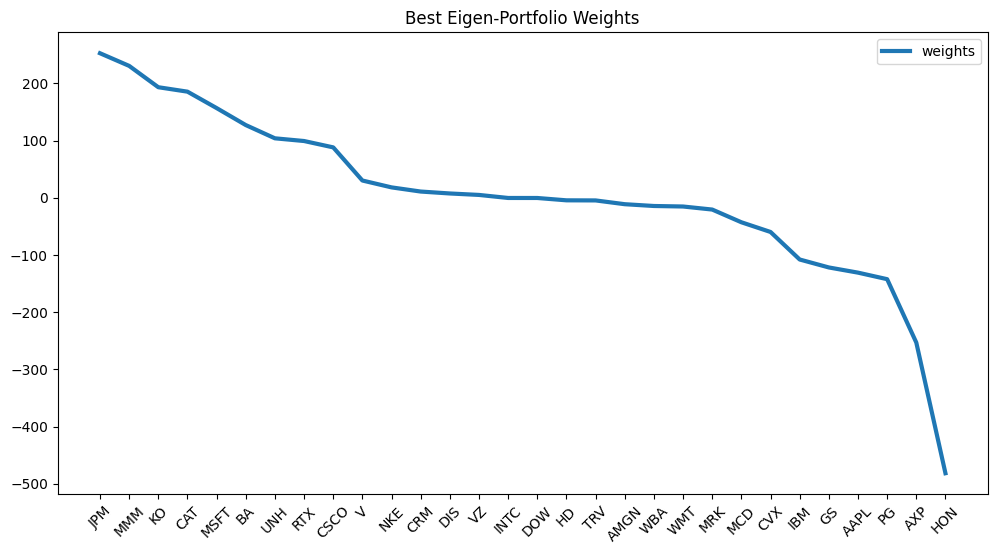}
    \caption{Portfolio weights of the optimal eigen-portfolio. Positive weights correspond to long positions, while negative weights indicate short positions.}
    \label{fig:best_eigen_weights}
\end{figure}

To further highlight the most influential positions, the table below presents the top five long and top five short positions by absolute weight. These stocks represent the largest directional exposures within the eigen-portfolio and may be interpreted as the primary drivers of its performance characteristics.

\begin{table}[H]
    \centering
    \begin{tabular}{lc}
        \hline
        \textbf{Stock Ticker} & \textbf{Weight(\%)} \\
        \hline
        JPM & 252.71 \\
        MMM & 230.60 \\
        KO  & 193.30 \\
        CAT & 185.56 \\
        MSFT & 156.86 \\
        GS	& -121.70 \\
        AAPL  & -130.76 \\
        PG	& -142.19 \\
        AXP	& -252.97 \\
        HON	& -481.37 \\
        \hline
    \end{tabular}
    \caption{Top five long and short positions in the optimal eigen-portfolio based on absolute weight magnitude, which sums up to 100\%}
    \label{tab:top_weights}
\end{table}

\subsection{Backtesting Performance}

To evaluate the out-of-sample performance of the optimal eigen-portfolio, we apply the corresponding portfolio weights—derived from the training set—to the test set returns. As a benchmark, we compare this strategy against a naive equally weighted portfolio, where each stock is assigned an identical weight.

The benchmark portfolio yields a Sharpe ratio of 0.99, with an annualized return of 10.86\% and annualized volatility of 11\%. These figures serve as a reference point for assessing the risk-adjusted effectiveness of the PCA-based strategy.

\begin{figure}[H]
    \centering
    \includegraphics[width=0.7\textwidth]{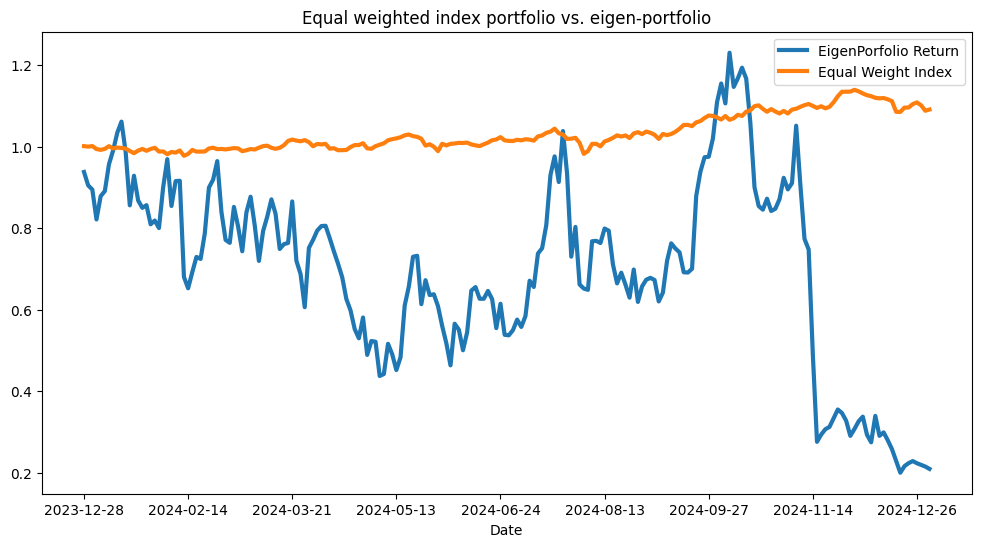}
    \caption{Comparison between the out-of-sample performance of the PCA-based eigen-portfolio and the equally weighted benchmark. The eigen-portfolio yields a Sharpe ratio of -0.56, with an annualized return of -85.47\% and volatility of 152.89\%, indicating significantly poorer performance relative to the benchmark.}
    \label{fig:comparison_1}
\end{figure}

\subsection{Diagnostic Analysis and Conclusion}

The out-of-sample underperformance of the PCA-based eigen-portfolio highlights several potential limitations of the strategy and invites further investigation:

\begin{itemize}
    \item \textbf{Overfitting to Training Data\cite{paikonen2007pca}:} The eigen-portfolio was selected based on its maximum Sharpe ratio within the training set. However, this selection process may overfit to noise or patterns that are not stable out-of-sample, especially if the Sharpe ratio was maximized over a large number of components.

    \item \textbf{Instability of Principal Components\cite{alibrahim2008stability}:} PCA is an unsupervised, data-driven method that is sensitive to the specific return structure in the training period. If the correlation matrix shifts in the test period, the principal components (and hence their associated weights) may no longer represent meaningful or profitable factors.

    \item \textbf{Market Regime Shift:} The test period may have experienced structural changes in market dynamics (e.g., volatility spike, macroeconomic shock) that made the training-based strategy ineffective.
\end{itemize}

It's also worth mentioning that when using a much less optimal weight from the training set (PC Index 5), we obtain a Sharpe ratio of 1.04, with annualized return being 57.36\% and volatility being 55.3\%. These results suggest that while PCA can reveal useful structure in return data, its application to portfolio construction requires careful validation, economic interpretation, and potentially additional techniques.

\section{Modification: Ensembling Top Eigen-Portfolios}

\subsection{Basic Idea}

The initial strategy of selecting a single eigen-portfolio—based solely on its in-sample Sharpe ratio—suffers from the risk of overfitting and instability. Principal components derived from PCA are sensitive to the specific statistical structure of the training data, and selecting only one component may lead to poor generalization in out-of-sample testing.

To address this, we propose an ensemble approach that combines multiple top-performing eigen-portfolios. Specifically, we identify the top-$N$ principal components based on their in-sample Sharpe ratios and construct a weighted combination of their associated portfolios. This method aims to diversify risk across multiple orthogonal return-driving directions, thereby improving robustness and stability.

\subsection{Mathematical Formulation}

Let $R \in \mathbb{R}^{T \times n}$ denote the standardized return matrix of $n$ assets over $T$ periods in the training set. Applying PCA to $R$ yields a set of $n$ orthonormal eigenvectors $\{ \mathbf{v}_1, \dots, \mathbf{v}_n \}$, which represent the principal directions in the asset return space.

Each eigenvector $\mathbf{v}_i \in \mathbb{R}^n$ can be interpreted as a portfolio weight vector (possibly rescaled) for the $i$-th eigen-portfolio. Let $P_i = R \mathbf{v}_i$ denote the return series of the $i$-th eigen-portfolio.

We compute the in-sample annualized Sharpe ratio $\mathcal{S}_i$ of each $P_i$ as:
\[
\mathcal{S}_i = \frac{\mu_i}{\sigma_i}, \quad \mu_i = \left( \prod_{t=1}^T (1 + r_{i,t}) \right)^{\frac{252}{T}} - 1, \quad \sigma_i = \text{Std}(P_i) \cdot \sqrt{252}
\]
where $r_{i,t}$ is the return of the $i$-th portfolio on day $t$, and 252 denotes the number of trading days in a year.

Let $\mathcal{I}_N$ be the index set of the top $N$ eigen-portfolios ranked by $\mathcal{S}_i$. The ensemble portfolio weight vector $\mathbf{w}_{\text{ens}}$ is then defined as:
\[
\mathbf{w}_{\text{ens}} = \sum_{i \in \mathcal{I}_N} \alpha_i \mathbf{v}_i, \quad \text{where} \quad \alpha_i = \frac{\mathcal{S}_i}{\sum_{j \in \mathcal{I}_N} \mathcal{S}_j}
\]

Remark: the sum of the indexes of $\mathbf{w}_{\text{ens}}$ is 1.

\subsection{Selecting the Optimal Number of Eigen-Portfolios for Ensembling}

To determine the optimal ensemble size \( N \)—i.e., the number of top eigen-portfolios to combine—we evaluate how the Sharpe ratio evolves as a function of \( N \) in the training set. 

\begin{figure}[H]
    \centering
    \includegraphics[width=0.7\textwidth]{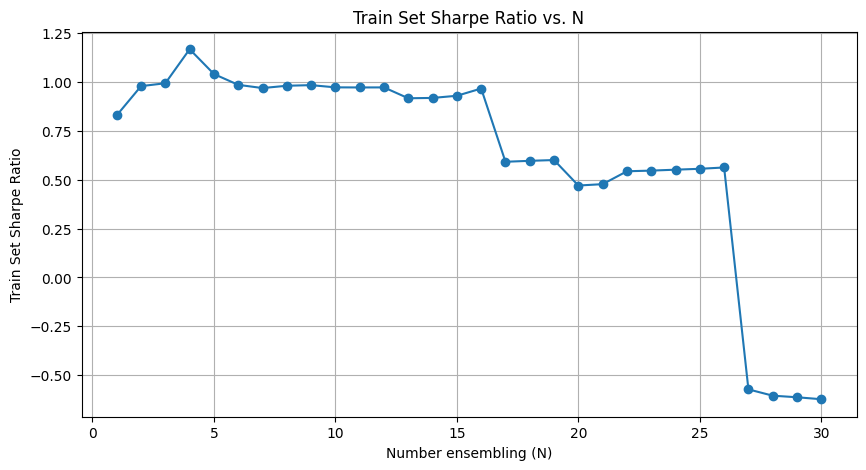}
    \caption{Sharpe ratio as a function of the number of eigen-portfolios used in the ensemble. The optimal value is achieved at \( N = 4 \), corresponding to a Sharpe ratio of 1.17.}
    \label{fig:ensemble_sharpe_vs_n}
\end{figure}

The figure above illustrates that the ensemble strategy achieves its best performance when combining the top 4 principal components, ranked by in-sample Sharpe ratio. Beyond this point, the marginal benefit of adding additional components diminishes or becomes negative, possibly due to the inclusion of noise-dominated eigenvectors.

\subsection{Resulting Portfolio Composition}

The portfolio weights of the ensemble corresponding to \( N = 4 \) can be extracted and interpreted as a blended exposure across the selected eigen-portfolios. For brevity, we display below the top five long and top five short positions (i.e., stocks with the largest positive and negative weights, respectively):

\begin{table}[H]
    \centering
    \begin{tabular}{lc}
        \hline
        \textbf{Stock Ticker} & \textbf{Weight (\%)} \\
        \hline
        TRV  & 10.86 \\
        CAT  & 10.57 \\
        WMT  & 6.71 \\
        CVX  & 4.86 \\
        AMGN & 3.99 \\
        MCD  & -5.22 \\
        PG   & -5.81 \\
        MMM  & -5.96 \\
        DOW  & -5.98 \\
        KO   & -6.96 \\
        \hline
    \end{tabular}
    \caption{Top five long and short positions in the ensemble portfolio constructed from the top 4 eigen-portfolios.}
    \label{tab:ensemble_top_weights}
\end{table}

\subsection{Backtesting Performance}

To evaluate the effectiveness of the ensemble approach, we apply the portfolio weights derived from the optimal ensemble (with \( N = 4 \)) to the test set. As in previous sections, we compare its performance against the equally weighted portfolio, which serves as a baseline benchmark.

\begin{figure}[H]
    \centering
    \includegraphics[width=0.7\textwidth]{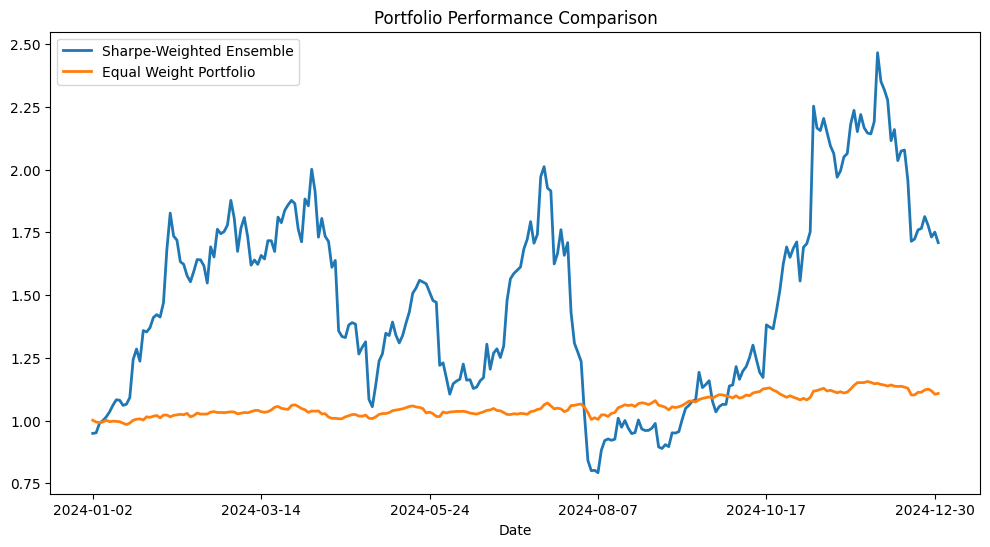}
    \caption{Out-of-sample performance comparison between the optimal ensemble portfolio and the equally weighted index portfolio. The ensemble portfolio achieves a Sharpe ratio of 1.05, with an annualized return of 93.79\% and volatility of 89.49\%.}
    \label{fig:ensemble_vs_equal}
\end{figure}

To provide a comprehensive comparison, Table~\ref{tab:three_portfolio_comparison} summarizes the out-of-sample performance metrics of three strategies:
\begin{itemize}
    \item The \textbf{equal-weight portfolio}, which allocates capital uniformly across all assets.
    \item The \textbf{best single-component portfolio}, derived from the single principal component with the highest in-sample Sharpe ratio.
    \item The \textbf{ensemble portfolio}, constructed by combining the top four eigen-portfolios as described in the previous section.
\end{itemize}

\begin{table}[H]
    \centering
    \begin{tabular}{l|r|r|r}
        \textbf{Portfolio Type} & \textbf{Annualized Return} & \textbf{Annualized Volatility} & \textbf{Sharpe Ratio} \\
        \hline
        Equal Weight          & 10.86\%  & 11.00\%   & 0.99 \\
        Single Component      & -85.74\% & 152.89\%  & -0.56 \\
        Best Ensemble (N=4)   & 93.79\%  & 89.49\%   & 1.05 \\
    \end{tabular}
    \caption{Out-of-sample comparison of three portfolio strategies: equal-weighted benchmark, single best eigen-portfolio, and top-$N$ ensemble approach.}
    \label{tab:three_portfolio_comparison}
\end{table}

The results show that the ensemble strategy significantly outperforms the single-component approach, both in terms of return and Sharpe ratio.

\section{Conclusion and Future work}

Our study has examined the application of Principal Component Analysis (PCA) in portfolio construction through the lens of eigen-portfolios. By decomposing the return covariance structure of Dow Jones Industrial Average (DJIA) constituents, we constructed orthogonal portfolios corresponding to principal components and evaluated their performance both in-sample and out-of-sample. While the single-component strategy—optimized based on in-sample Sharpe ratio—demonstrated strong historical performance, its substantial deterioration in the test set highlights the risk of overfitting and the inherent instability of PCA when applied naively.

To address these limitations, we proposed an ensemble methodology that aggregates the top-N eigen-portfolios ranked by in-sample Sharpe ratio. This ensemble approach significantly improved out-of-sample performance, achieving a higher Sharpe ratio than both the single-component strategy and the equal-weighted benchmark. These findings underscore the value of diversification not only across assets, but also across latent risk factors derived from statistical decomposition. 

Thus, the ensemble eigen-portfolio approach offers a relatively straightforward yet effective method for portfolio construction, particularly for investors seeking to outperform standard benchmarks or achieve superior market-relative returns. Nonetheless, several important avenues remain open for future research:

\begin{itemize}
    \item \textbf{Robustness and Stability:} Traditional PCA is sensitive to estimation error and outliers. Future work may explore robust PCA techniques, including shrinkage estimators and sparse PCA, to enhance stability in the presence of noise and high dimensionality which are introduced in \cite{he2023robustPCA} and \cite{cai2021learnedRPCA}.
    \item \textbf{Time-Varying Structures:} The correlation structures underpinning PCA are inherently dynamic. Incorporating rolling-window PCA or dynamic factor models could allow the strategy to adapt to regime shifts and evolving market conditions \cite{alshammri2021mdpca}.
     \item \textbf{Nonlinear structure:} PCA is inherently linear and may fail to capture complex structures in financial data. Nonlinear generalizations such as kernel PCA, autoencoders, or manifold learning techniques offer promising directions for capturing richer latent representations\cite{allerbo2021pathlassoAE}.
\end{itemize}

\pagebreak

\setcounter{section}{0}

\renewcommand{\thesection}{\Alph{section}} 

\newcommand{\pythonstyle}{
  \lstset{
    language=Python,
    basicstyle=\ttfamily\small,
    keywordstyle=\color{blue},
    commentstyle=\color{gray},
    stringstyle=\color{red},
    showstringspaces=false,
    numbers=left,
    numberstyle=\tiny,
    breaklines=true
  }
}

\lstnewenvironment{python}[1][]{
  \pythonstyle
  \lstset{#1}
}{}

\end{document}